\documentclass[preprint,aps,prl,showpacs,amsmath,amssymb]{revtex4}
\usepackage{graphics}
\usepackage{epsf,graphicx,epsfig,psfrag}
\begin{document}

\title{Magnetothermoelectric effects in Fe$_{1+d}$Te$_{1-x}$Se$_x$}
\author{Marcin Matusiak$^1$, Ekaterina Pomjakushina$^2$ and Kazimierz Conder$^2$}
\affiliation{1.Institute of Low Temperature and Structural
Research, Polish Academy of Sciences, P.O. Box 1410, 50-950 Wroclaw, Poland}
\affiliation{2. Laboratory for Developments and Methods, Paul Scherrer Institute, CH-5232 Villigen, Switzerland} 
\date{\today}

\begin{abstract}
We report resistivity as well as the Hall, Seebeck and Nernst coefficients data for Fe$_{1+d}$Te$_{1-x}$Se$_x$ single crystals with $x$ = 0, 0.38, and 0.40. In the parent compound $\rm Fe_{1.04}Te$ we observe at $T_N$ = 61 K a sudden change of all quantities studied, which can be ascribed to the Fermi surface reconstruction due to onset of the antiferromagnetic order. Two very closely doped samples: $\rm Fe_{1.01}Te_{0.62}Se_{0.38}$ (Se38) and $\rm Fe_{1.01}Te_{0.60}Se_{0.40}$ (Se40) are superconductors with $T_c$ = 13.4 K and 13.9 K, respectively. There are no evident magnetic transitions in either Se38 or Se40. Properties of these two single crystals are almost identical at high temperatures, but start to diverge below $T \approx$ 80 K. Perhaps we see the onset of scattering that might be a related to  changes in short range magnetic correlations caused by selenium doping.

\end{abstract}


\maketitle
\section{1. Introduction}
The discovery of the simplest iron-based superconductor, the chalcogenide Fe$_{1+d}$Te$_{1-x}$Se$_x$ \cite{Hsu,Yeh}, gives hope for understanding the pairing mechanism in the entire family of new Fe-based materials. Yet soon it become apparent that the ground state of the parent compounds of the iron-chalcogenide and iron-pnictides were magnetic phases of a different nature. Namely, the antiferromagnetic (AFM) spin-density-wave (SDW) order observed in the "1111" \cite{Dong,Matusiak} and "122" families \cite{Rotter,Matusiak2} seems to be absent from "11" Fe-chalcogenides, despite these compounds have similar electronic band structure \cite{Xia}. The in-plane magnetic wave vector of the AFM order in the Fe$_{1+d}$Te parent compound has been identified to be ($\pi$,0) \cite{Bao,Xia}, whereas SDW in iron pnictides forms with the nesting vector ($\pi$,$\pi$) \cite{Chen1}. This discrepancy raised a question whether superconductivity in "11" has the same origin as in other iron-based families \cite{Xia}. Nonetheless, the recent inelastic neutron scattering measurements shows that ($\pi$,$\pi$) spin correlations, which are common for iron-pnictides \cite{Cruz,Huang} and copper-based superconductors \cite{Cheong,Mook}, develops in Fe$_{1+d}$Te$_{1-x}$Se$_x$ upon selenium doping \cite{Lumsden,Liu}.

In this paper we take a closer look at the region of the phase diagram of Fe$_{1+d}$Te$_{1-x}$Se$_x$, where the ($\pi$,$\pi$) spin fluctuations are expected to overwhelm ($\pi$,0) ones. To this end we study transport properties of parent iron chalcogenide $\rm Fe_{1.04}Te$ as well as two selenium doped superconductors: $\rm Fe_{1.01}Te_{0.62}Se_{0.38}$ (Se38) and $\rm Fe_{1.01}Te_{0.62}Se_{0.40}$ (Se40). The results show rather sudden change of the low temperature behavior, which may be related to moving (upon Se-doping) from region of coexistence of ($\pi$,$\pi$) and ($\pi$,0) fluctuations, to the region of domination of ($\pi$,$\pi$) fluctuations. In such a case it is likely that different types of spin fluctuations interact with different conducting bands. 

\section{2. Material and methods}
Single crystals of $\rm Fe_{1.01}Te_{0.6}Se_{0.4}$ and $\rm Fe_{1.04}Te$  nominal compositions were grown using a modified Bridgeman method. The respective amounts of Fe, Se and Te powders of a minimum purity of 99.99\% were mixed and pressed into a rod of 7 mm diameter and placed into evacuated sealed quartz ampoule. This ampoule was sealed into a second quartz ampoule. The rod was first melted and homogenized at 1200 $\rm ^o$C for 4 hours, cooled in a temperature gradient at a rate 4 $\rm ^o$C/h down to 750 $\rm ^o$C followed by 50 $\rm ^o$C/h cooling \cite{Khasanov}. For this study we cut out one single crystal from the $\rm Fe_{1.04}Te$ rod and two single crystals from different places of the $\rm Fe_{1.01}Te_{0.6}Se_{0.4}$ rod. The chemical compositions of these two Se-doped crystals were determined by the energy dispersive x-ray (EDX) analysis, and they turned out to differ slightly in Se-content ($x$ = 0.38 and 0.40). The data were collected in four independent places for both samples and the noticed variation in composition (a difference between the highest and lowest reading) was smaller than one percent for every element.

The resistivity was measured using the four-probe technique with 25 $\mu$m gold wires attached to the sample with two component silver epoxy (EPO-TEK H20E). For the Hall coefficient measurement, the sample was mounted on a rotatable probe and continuously turned by 180 degree (face down and up) in a magnetic field ($B$) of 12.5 T to effectively reverse the field anti-symmetrical signal. During the thermoelectric power and Nernst coefficient measurements, the sample was clamped between two phosphor bronze blocks, which had two Cernox thermometers and resistive heaters attached to them. The temperature runs were performed in magnetic fields from -12.5 to +12.5 T in order to extract the field voltage components that were odd and even in $B$.

\section{3. Results}
The Fe$_{1+d}$Te parent compound is known to have the first-order magneto-structural transition around $T_N \approx 60$ K \cite{Li}. This transition strongly affects all transport coefficients studied here, as shown in Figure 1.
\begin{figure}
\label{Fig1}
 \epsfxsize=9cm \epsfbox{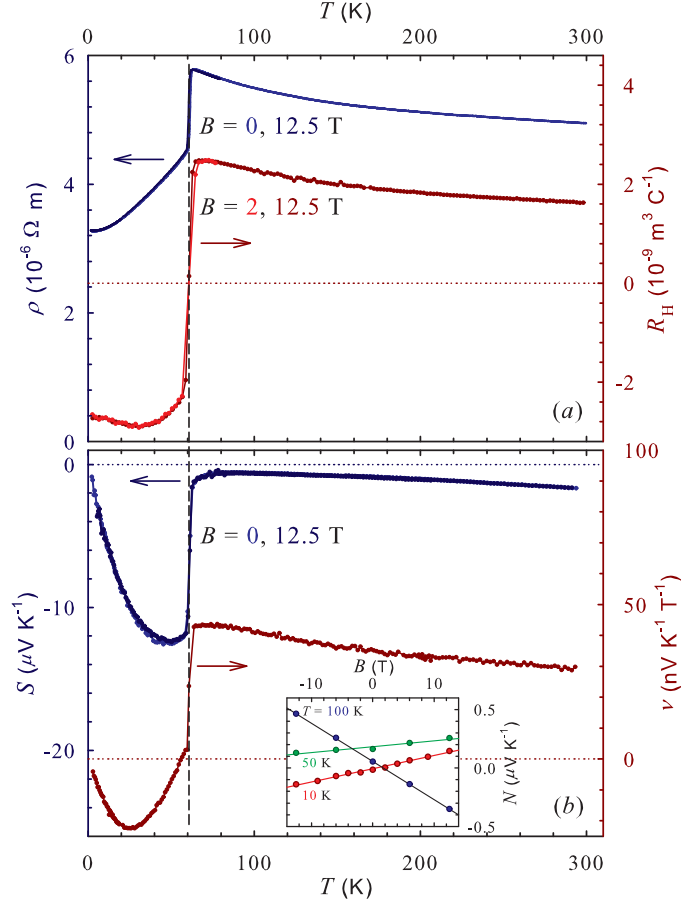}
 \caption{(Color online) The temperature dependences of the resistivity and Hall effect (upper panel a) as well as the Seebeck and Nernst effects (bottom panel b) of the $\rm Fe_{1.04}Te$ single crystal. The dashed line marks the structural/magnetic transition at $T_N$ = 61 K. Figure presents data taken for the maximal magnetic field of $B$ = 12.5 T (dark lines) as well for smaller (or absent) fields (light lines), but the difference is insignificant. The inset in panel b shows field dependence of the Nernst signal, which also does not show deviation from linearity.}
 \end{figure}
The step-like change in the temperature dependences of the electrical resistivity ($ \rho$), as well as Hall ($R_H$) coefficient at $T_N$ was suggested to occur due to reconstruction of the Fermi surface due to formation of the antiferromagnetic order \cite{Chen}. It seems that this reconstruction can be related to formation of spin stripes, which are expected to produce a maximum (or minimum) in the Nernst signal at $T \approx 1/3 T_{N}$ \cite{Hackl}. In fact, such a minimum is observed in $ \nu(T)$ of $\rm Fe_{1.04}Te$ at $T = 25 K$.

As reported previously, there is practically no detectable influence of the magnetic field up to $B =$ 12.5 T on the resistivity \cite{Chen}, thermoelectric power \cite{Pallecchi} and Hall coefficient \cite{Chen}. The Nernst effect appears to be no exception here. The magnetic field dependence of the Nernst signal ($N(B)$) shows no deviation from linearity in the temperature and magnetic field studied (see inset in Fig. 1b). A lack of response of the transport properties to the magnetic field in $\rm Fe_{1.04}Te$ suggests strong magnetic coupling, which is characterized by the in-plane wave-vector ($\pi$,0) \cite{Bao,Xia}. Nonetheless, the antiferromagnetic ($\pi$,0) order in Fe$_{1+d}$Te$_{1-x}$ can be suppressed by substitution of tellurium with selenium \cite{Khasanov}. Moreover, such a substitution introduces ($\pi$,$\pi$) fluctuations that seem to be directly related to formation of the superconductivity \cite{Lumsden,Liu}, whereas ($\pi$,0) fluctuations were attributed to low temperature weak charge carrier localization effects \cite{Liu}.

In order to study these phenomena we measured two very closely substituted single crystals: Fe$_{1.01}$Te$_{0.62}$Se$_{0.38}$ (Se38) and Fe$_{1.01}$Te$_{0.60}$Se$_{0.40}$ (Se40). The critical temperatures of these samples, defined as the maximum in d$\rho$/d$T$, are  similar: $T_c$ = 13.4 K for Se38 and $T_c$ = 13.9 K for Se40. The resistivity of Se38, shown in Fig. 2a, rises below $T_{loc} \approx$ 40 K, which can be a result of the aforementioned charge carrier localization. The presented in Fig. 2a $R_H(T)$ of the same sample rises with decreasing $T$ for the entire temperature range studied, but this trend becomes steeper below $T_{loc}$. The behavior of the $\rho(T)$ and $R_H(T)$ below  $T_{loc}$ might be related, since in the simple one-band picture: $\rho \propto n^{-1}$ as well as $R_H \propto n^{-1}$ ($n$ is the charge carrier concentration). The field dependence of the Hall coefficient observed below $T_{loc}$ could be understand within the same picture. Namely, if the observed localization is due to the onset of the short range AF correlations, the magnetic field will decrease this effect by quenching fluctuations. As discussed later, this may be better seen in Hall effect than in the resistivity in a case of a multiband conductor, and Fe$_{1+d}$Te$_{1-x}$Se$_x$ is known to have a complex  electronic band structure \cite{Chen1}.

\begin{figure}
\label{Fig2}
 \epsfxsize=9cm \epsfbox{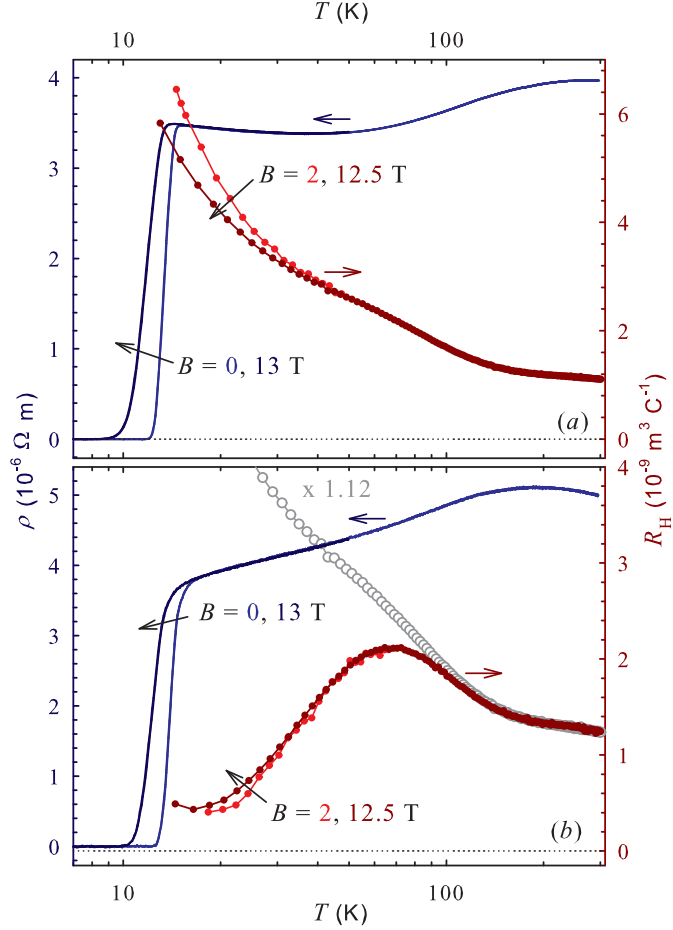}
 \caption{(Color online) The temperature dependences of the resistivity and Hall coefficient of the Fe$_{1.01}$Te$_{0.62}$Se$_{0.38}$ (panel a) and  Fe$_{1.01}$Te$_{0.60}$Se$_{0.40}$ (panel b) single crystals. The gray open points in the panel b depict $R_H(T)$ for Se38 normalized to the room temperature value of $R_H(T)$ for Se40.}
 \end{figure}

Figure 2b presents the resistivity and Hall coefficient for the second doped crystal: Se40. The high temperature dependence of both coefficients are very similar to those for Se38, but their low temperature parts differ obviously for these closely doped samples. Namely, there is no sign of localization in $\rho(T)$, and $R_H(T)$ start to decreases below $T \approx$ 40 K. The $B$-dependence of $R_H$ is weaker in Se40 than it was in Se38, and for Se40 application of the magnetic field causes a increase of the Hall coefficient at low temperatures. The qualitative difference between the low temperature parts of $\rho(T)$ and $R_H(T)$ in samples with close stoichiometry may suggest that we observe a sudden crossover from region of domination of ($\pi$,0) magnetic order fluctuations (which are antagonistic to metallicity) to region of domination of the ($\pi$,$\pi$) fluctuations (which are supposed to promote superconductivity).
 It is important to emphasize \cite{Liu2} that both Fe$_{1+d}$Te$_{1-x}$Se$_x$ single crystals have the same iron content as determined by EDX analysis.

Surprisingly the normal state temperature dependences of the thermoelectric power and Nernst coefficient for crystals, which have substantially different low temperature $\rho(T)$ and $R_H(T)$, look similar in the entire temperature range. $S(T)$ for Se38 and Se40 presented in the semi-logarithmic scale in Fig. 3a saturates at high temperatures to a small positive value, whereas at low temperature change sign and exhibit negative minimum, which presence was previously reported \cite{Pallecchi}. The thermoelectric power is independent of $B$ up to 12.5 T, with exception for the region near superconducting transition, where existence of the normal state is extended by application of the magnetic field. The minimum in $S(T)$ is unlikely to be of the phonon-drag origin, since in such a case a size of this minimum is expected to be field dependent \cite{Blatt,Gurevich}.
\begin{figure}
\label{Fig3}
 \epsfxsize=10cm \epsfbox{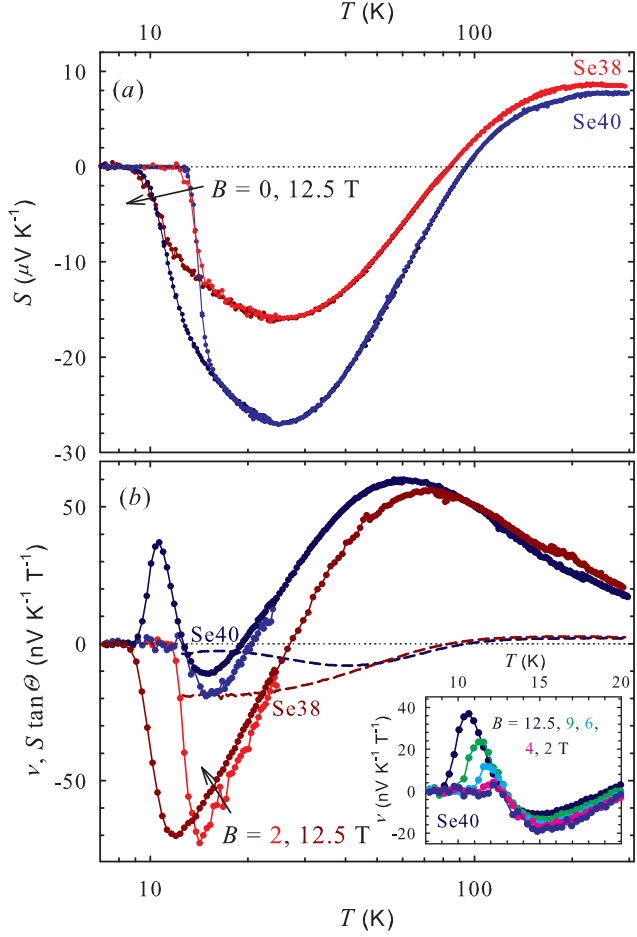}
 \caption{(Color online) The temperature dependences of the Seebeck (upper panel a) and Nernst (bottom panel b) coefficients of the the Fe$_{1.01}$Te$_{0.62}$Se$_{0.38}$ (Se38) and  Fe$_{1.01}$Te$_{0.60}$Se$_{0.40}$ (Se40) single crystals. The dashed lines in panel b depict $S \tan\theta$ terms for Se38 and Se40 ($B$ = 12.5 T). Inset in panel b shows the temperature dependences of the Nernst coefficient of Se40 in the vicinity of $T_c$ measured at various $B$.}
 \end{figure} 
Analogously, there is no substantial difference between the normal state Nernst effect for Se38 and Se40, which are shown Figure 3b in the semi-logarithmic scale. Values of $\nu$ for both Se38 and Se40 are positive at the room temperature, rise slowly upon decreasing temperature and exhibit sudden drop below $T \approx$ 70 K for Se38 and $T \approx$ 60 K for Se40. We do not observe any contribution from the vortex motion to the Nernst effect in Se38, while in Se40 there is a positive peak at $T_c$, which has been identified as a flux flow effect \cite{Pourret}. The magnetic field dependence of $\nu$ in the normal state of Se38 and Se40 is small and limited to temperatures below $T \approx$ 25 K. This field dependent signal cannot appear due to superconducting fluctuation, because in such a case $\nu$ would decrease with magnetic field \cite{Wang}. The field dependence of the Nernst coefficient has likely the same origin as supposed for the $B$-dependence of the Hall coefficient, i.e. suppression of the AF fluctuations by the magnetic field. Noticeably, the applied magnetic field causes decreasing of in $\nu$ in both samples, whereas influence of $B$ on $R_H$ is opposite in Se38 and Se40. It might mean that in the Hall data we see suppression of different type of fluctuations, while the temperature dependences of the Nernst effect are governed by the same type of the short range magnetic correlations.

The normal-state Nernst coefficient is composed of two terms: $\nu = (\rho \alpha_{xy} - S \tan \theta)/B$, where $\alpha_{xy}$ stands for the off-diagonal element of the Peltier tensor, and $\tan \theta$ is the Hall angle \cite{Wang2}. In simple metals these two terms are comparable and in consequence $\nu$ is very small \cite{Sondheimer}. In Se38 and Se40 $\nu >> (S \tan \theta)/B$ (see Fig. 3b) with the exception for the temperatures close to the point, where the Nernst coefficient changes its sign.

\section{4. Discussion}
We focus on analyzing data from two selenium doped single crystals: Fe$_{1.01}$Te$_{0.62}$Se$_{0.38}$ and  Fe$_{1.01}$Te$_{0.60}$Se$_{0.40}$. As expected for such closely doped samples all four transport coefficients measured for Se38 look almost identical to those for Se40, but only at the high temperatures. This suggests that differences between Se-doped samples regard a mechanism that is limited to low temperatures. One of possibilities is that at low-$T$ we can see an influence of short range spin fluctuations, which are known to change their character in Fe$_{1+d}$Te$_{1-x}$Se$_x$ upon Se-doping. Namely, the ($\pi$,0) fluctuations, which dominate in samples with low selenium content, are replaced by ($\pi$,$\pi$) for higher $x$ \cite{Liu}. The ($\pi$,0) fluctuations are believed to be responsible for weak charge carrier localization noticeable in the low temperature part of $\rho(T)$ and $R_H(T)$ dependences of Se38. This sample has a sharp superconducting transition, but at the same time it is anticipated to be a non-bulk superconductor \cite{Liu}. Measurements of the magnetization presented in Fig. 4 support these expectations.
\begin{figure}
\label{Fig4}
 \epsfxsize=10cm \epsfbox{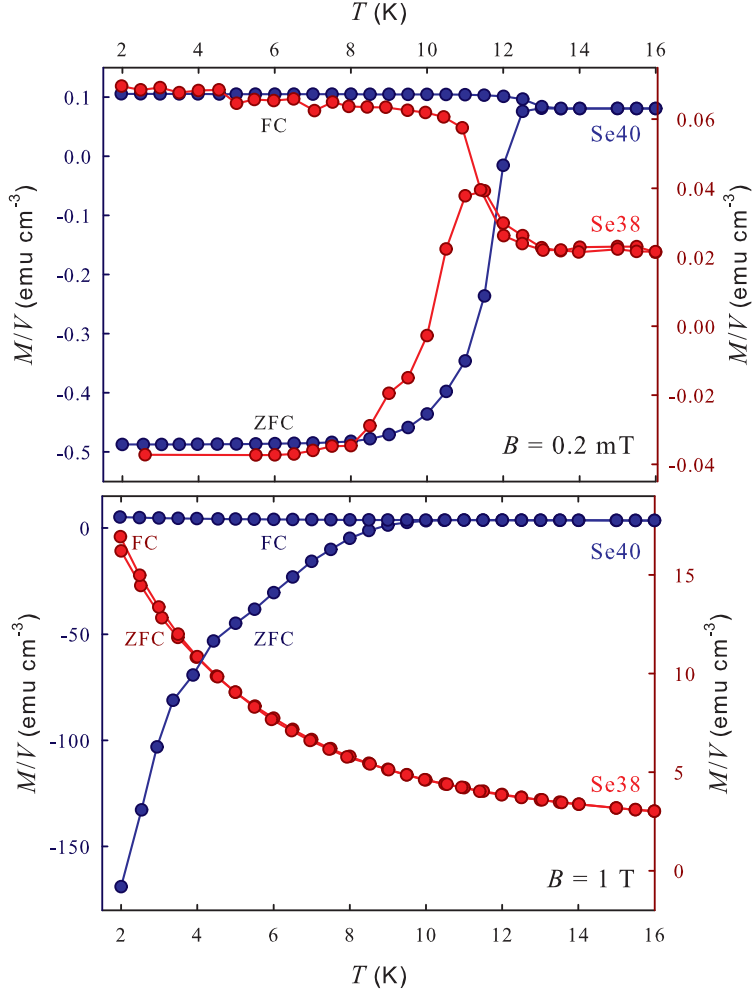}
 \caption{(Color online) The temperature dependences of the magnetization for  the Fe$_{1.01}$Te$_{0.62}$Se$_{0.38}$ (Se38) and  Fe$_{1.01}$Te$_{0.60}$Se$_{0.40}$ (Se40) single crystals measured at $B$ = 2 mT (upper panel) and $B$ = 2 T (bottom panel). Acronims ZFC and FC refer to "zero-field-cooled" and "field-cooled" runs, respectively.}
 \end{figure} 
In very low magnetic field of $B$ = 2 mT both Se38 and Se40 cooled in zero field develop a  diamagnetic signal below $T_c^{mag}$ = 11.5 and 12.5 K, respectively. In the field of $B$ = 1 T Se40 still shows superconducting transition at $T_c^{mag}$ = 10 K, while there is practically no detectable diamagnetic signal in Se38 despite the electrical resistivity in this sample drops to zero at $T \approx$ 11.5 K. This observation stays in agreement with the presented here Nernst data. Namely, there is no Nernst signal from the vortex flow near the superconducting transition in Se38, because vortices in a mixed state that occupies only a small volume fraction of the sample are not able to produce sizable transverse electric signal. In the contrary, we found a clear signal from vortex liquid in the Se40 crystal. Untypically, the size of peak in $\nu(T)$ at $T_c$, presented in the inset in Fig. 3b, rises with the increasing magnetic field and become almost undetectable for $B$ = 1 T. Perhaps this is due to wide region of existence of the vortex solid phase, where "frozen" vortices do not contribute to the Nernst signal.
An important question is why the low temperature part of the Seebeck and Nernst effects are weakly affected by change of type of the dominant short range magnetic correlations, whereas differences in $\rho(T)$ and $R_H(T)$ are very pronounced. There are, basically, two possible explanations, which do not rule out each other. One is that the Nernst and Seebeck effects are strongly influenced by the shift of the chemical potential which overwhelm consequences of variation of the relaxation times. The other is that the thermoelectric transport in Fe$_{1+d}$Te$_{1-x}$Se$_x$ is dominated by a conductive band, which properties are mainly affected by only one type of fluctuations. There are arguments to support each of these scenarios, but both effects may occur simultaneously.

In the relaxation time approximation the electrical conductivity ($\sigma$) is simply proportional to the relaxation time ($\tau$): $\sigma = \frac{n e^2 \tau}{m^*}$, where $e$ is the elementary charge, and $m^*$ is the effective mass. In a multiband system the total (index $t$) electrical conductivity is a sum of conductivities from different bands (index $i$):
\begin{equation}
\sigma_t = \sum \sigma_i .
\end{equation}
The Hall coefficient of a given band does not depend directly on the relaxation time ($R_{H,i} = \frac{1}{n_i e}$), but in a multiband conductor total $R_H$ is a sum of partial contributions weighted by square of $\sigma_i$:
\begin{equation}
R_{H,t} = \frac {\sum R_{H,i} \sigma_i^2}{\sigma_t^2} .
\end{equation}
This make $R_{H,t}$ even more sensitive to changes in the relaxation times than $\sigma_t$, and data presented in Fig. 1 stay in agreement with this conclusion.

One can expect a similar, but weaker effect in the thermoelectric power, which is also a sum of contributions from different bands, where $S_i$ are weighted with conductivities without squares:
\begin{equation}
S_t = \frac{\sum S_i \sigma_{i}}{\sigma_{t}} .
\end{equation}
$S_i$ in the Mott-Jones formula is not simply related to the value of $\tau$, but rather to the logarithmic derivative of $\tau$ with energy ($\varepsilon$): $S_i = - \frac{\pi^2 k_B^2 T}{3|e|} (\frac{1}{\sigma_i} \frac{\partial \sigma_i}{\partial \varepsilon})_{\varepsilon = E_F}$. On the other hand, Fe$_{1+d}$Te$_{1-x}$Se$_x$ is supposed to be a metal with the extremely small Fermi energy ($E_F$) \cite{Lubashevsky}, and in this case the chemical potential ($\mu$) can be temperature dependent. This  influences strongly the thermopower, which is by definition related to variation of the electrochemical ($\overline{\mu}$) rather than only electrical ($\psi$) potential:
\begin{equation}
S \equiv -\frac{\nabla \overline{\mu} }{e \nabla T} ,
\end{equation}
where $\overline{\mu} = \mu + e \psi$.
In a system with low carrier concentration, which is the case of Fe$_{1+d}$Te$_{1-x}$Se$_x$ \cite{Pourret}, changes of the chemical potential can dominate the $S(T)$ dependence, and at the same time one can expect a saturation of the Seebeck coefficient in the high temperature limit \cite{Durczewski} that is in fact observed in both Se38 and Se40 (Fig. 3).

A value of the Nernst coefficient is in the zero-temperature limit directly connected with $\varepsilon_F^{-1}$: $\frac{\nu}{T} = -\frac{\pi^2 k_B}{3e} \frac{T\mu_H}{\varepsilon_F}$ \cite{Behnia}, thus one may expect similar relation of $\nu$ to $\mu$ in finite temperatures. However, the case of the Nernst effect in a multiband system is more complicated, since in the presence of positive (index $p$) and negative (index $n$) charge carriers, there is additional contribution to the Nernst signal from ambipolar flow of quasiparticles \cite{Sondheimer}:
\begin{equation}
\nu = \frac{\nu_p \sigma_p + \nu_n \sigma_n}{\sigma_p + \sigma_n} + \frac{\sigma_p \sigma_n (S_p - S_n) (R_{Hp} \sigma_p - R_{Hn} \sigma_n)}{(\sigma_p + \sigma_n)^2},
\end{equation}

The aforementioned enhancement of $\nu$ over $(S \tan\theta)/B$ at high temperatures may be related to the ambipolar term \cite{Bel}, but at low temperatures the mechanism must be different, since the contribution from the ambipolar flow is always positive \cite{Delves}, whereas the sudden drop of the Nernst coefficient below $T \approx 60$ K that causes change of the $\nu$ sign to negative. This indicates that Nernst signal is, at least at low temperatures, dominated by one of conductivity bands. A similar conclusion was drawn on the basis of thermoelectric studies by Pourret et al., who suggested that the superconducting and normal properties of Fe$_{1+y}$Te$_{0.60}$Se$_{0.40}$ are dominated by a single electron-like band \cite{Pourret}. Fe$_{1+y}$Te$_{0.60}$Se$_{0.40}$ has been also identified as a metal with strong electronic correlations. In such a case the temperature dependence of thermoelectric power can be far from simply linear one \cite{Arsenault}. We believe that the presented results can be better understood in the light of inelastic neutron scattering measurements showing that the ($\pi$,$\pi$) and ($\pi$,0) spin fluctuations coexist over a wide composition range, but the latter are strongly suppressed in the region of occurrence of bulk superconductivity \cite{Liu}. This leads to conclusion the resistivity and Hall effect show the sudden suppression of the ($\pi$,0) fluctuations, whereas the Nernst and Seebeck effects are dominated by a band, which is affected mainly by the ($\pi$,$\pi$) fluctuations that vary slowly with Se-doping. This means that different conducting bands are differently coupled to the spin fluctuations which compete with or promote superconductivity. Therefore, the band, which dominates thermoelectric transport, is likely to be primarily responsible for superconductivity.

\section{5. Conclusions}
The respective properties of two closely doped single crystals Fe$_{1.01}$Te$_{0.62}$Se$_{0.38}$ and Fe$_{1.01}$Te$_{0.60}$Se$_{0.40}$ are almost identical at high temperatures, but they start to diverge below $T \approx$ 80 K. Both, Se38 and Se40, samples are superconducting, but the Nernst and magnetization data suggest that bulk superconductivity is absent from Se38, and present in Se40. Differences in the normal-state resistivities and Hall coefficients are pronounced, whereas the temperature dependences of the Nernst and Seebeck coefficients remain slightly affected. We conclude that this could be a consequence of the sudden suppression of the ($\pi$,0) magnetic correlations by Se-doping. In such a case, the low temperature thermoelectric properties would be dominated by one band, which could be a host for superconductivity. However, we cannot definitely exclude a possibility that we see the onset of a scattering mechanism of different origin, whereas the Nernst and Seebeck effects are governed by the temperature shift of the chemical potential.

\section*{Acknowledgments}
The authors are grateful to K. Durczewski, K. Rogacki and J.R. Cooper for useful comments and to A.M. Krzton-Maziopa and M. Malecka for performing the EDX analysis. This work was supported by a grant No. N N202 130739 of the Polish Ministry of Science and Higher Education and NCCR MaNEP.


\end{document}